\documentclass{article}

\usepackage{arxiv}

\usepackage[utf8]{inputenc} 
\usepackage[T1]{fontenc}    
\usepackage{hyperref}       
\usepackage{url}            
\usepackage{booktabs}       
\usepackage{amsfonts}       
\usepackage{nicefrac}       
\usepackage{microtype}      
\usepackage{graphicx}
\usepackage{siunitx}
\usepackage[numbers,sort&compress]{natbib}
\usepackage{doi}

\title{The SWaP plot: Visualising the performance of portable atomic clocks as a function of their size, weight and power.}

\usepackage{authblk}

\author[1]{Benjamin R. White\thanks{Corresponding author: \texttt{benjamin.white@adelaide.edu.au}}}
\author[2]{Rachel F. Offer}
\author[1]{Ashby P. Hilton}
\author[1]{Andre N. Luiten}

\affil[1]{Institute for Photonics, Advanced Sensing and Quantum Technology (IPAS-QT), Adelaide University, Adelaide, Australia 5005}
\affil[2]{Department of Physics, University of Strathclyde, Glasgow, G40 NG, United Kingdom}

\date{}

\renewcommand{\headeright}{Technical Report. August 2026}
\renewcommand{\undertitle}{Technical Report. Updated August 2026}

\renewcommand{\shorttitle}{The SWaP plot}

\begin{document}
\maketitle

\begin{abstract}
    Precision timekeeping is fundamental to modern technologies such as Global Navigation Satellite Systems (GNSS), communication networks, financial transactions, and power grid management. Over the past 50 years, microwave atomic clocks have been the standard for timing precision. The new generation of optical atomic clocks have demonstrated orders of magnitude better performance, and are now transitioning from research to practical applications. We provide a web resource that tracks the performance of these optical atomic clocks, measured in terms of their Allan deviation at various integration times, against their SWaP requirements via an interactive plot. The most current data and additional resources are available online, providing a continuously updated reference for those interested in precision timing.
\end{abstract}

\keywords{Precision timekeeping \and Optical atomic clocks \and Microwave atomic clocks \and Portable atomic clocks \and SWaP (Size, Weight, and Power) \and Allan Deviation (ADEV) \and Clock performance
}

\section{Motivation}
    As optical atomic clocks transition from research labs to practical applications, we have developed a website which aims to serve as a resource tracking this progression as it happens. The
    URL links are provided in section ~\ref{sec:iLinks} of this document.
    
    Precision timekeeping underpins most modern technologies, including navigation, communication network synchronization, financial transactions, and power grid management, among many others. Historically, advancements in precise timekeeping have paralleled nearly every technological revolution. Over the last 50 years, microwave atomic clocks, now essential to our infrastructure, have set the standard for atomic timing precision.
    
    The main focus is on the immediate challenges to deploying portable atomic clocks, namely, their SWaP requirements, and how these critical factors directly influence clock performance and deployability. Beyond this, there will be other considerations relating to the ability to manufacture at scale, cost, accuracy, reliability and other aspects of how clock specifications might meet various use-cases. However, with this document we focus on the relationship between SWaP and frequency stability.

    We plan to update this report periodically such that the reference list provided herein remains up to date as new systems are published and or announced. However the website will remain the most frequently updated data source.

\begin{figure}
    \centering
    \includegraphics[width=1\linewidth]{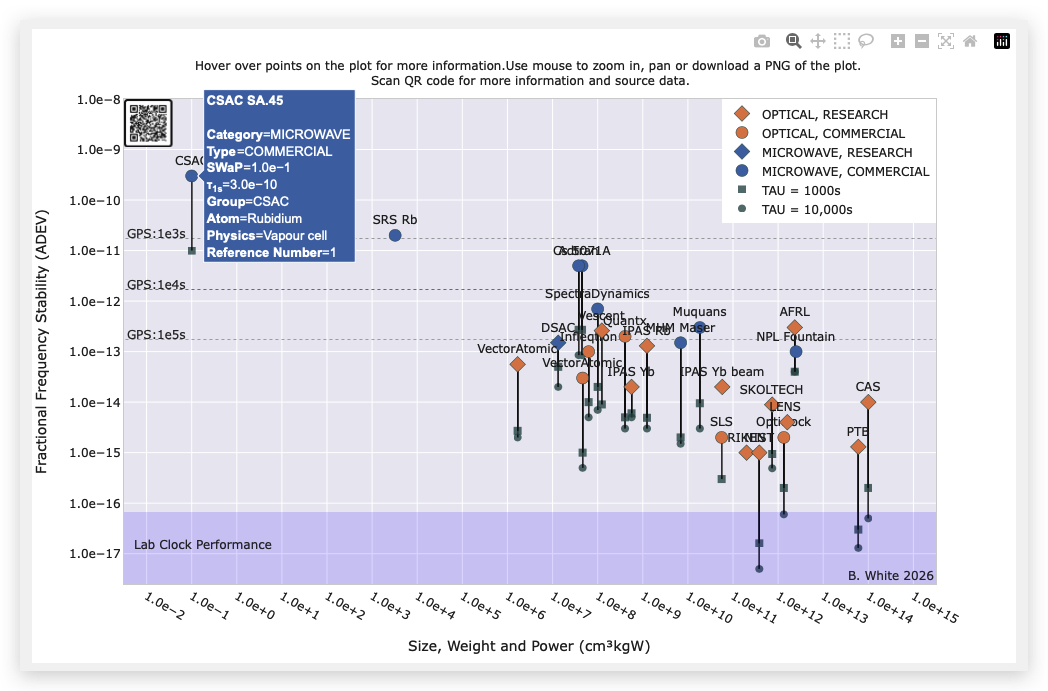}
    \caption{
        Static screenshot of the current (August 2026) version of the SWaP plot.
        The horizontal axis represents the combined SWaP of the device in units of \qty{}{\cm^3\kg\W}. The vertical axis is the fractional frequency instability. Where available, a device will have an ADEV value for different integration times, connected by thin vertical lines (see plot legend). The dashed horizontal line is indicative of GNSS timing performance at 1 second. The online versions of this plot are interactive,  allowing the user to zoom in on congested areas of the plot and hovering over a point to view more details for each device (shown for the CSAC SA.45). There is a menu visible at the very top right of the plot that will appear when hovering over that region. This allows the user to download a PNG image of the plot, zoom, pan and perform other interactive actions. Selected references available for spec sheets and published works \cite{CSAC2024, 
        PRS102024, 
        5071A2024, 
        MHMMaser2024, 
        NPLCsFountain2024, 
        MUCLOCK2024, 
        cRb2024, 
        VECTORATOMIC2024, 
        TIQKER2024, 
        OPTICLOCK2024, 
        DFMVESCENT2024, 
        Burt2021,
        Huang2020,
        Koller2017,
        Poli2014,
        Hilton2025,
        Takamoto2020a, 
        Khabarova2022,
        AdtranOSA33002024,
        SLS2026,
        Schioppo2017,
        Bothwell2025, 
        Offer2026}.
        For complete list, see the current online version. Scanning the QR code will link to the IPAS-QT precision timing plot page.}
    \label{fig:FULL-SWAP-PLOT}
\end{figure}

\begin{figure}
    \centering
    \includegraphics[width=1\linewidth]{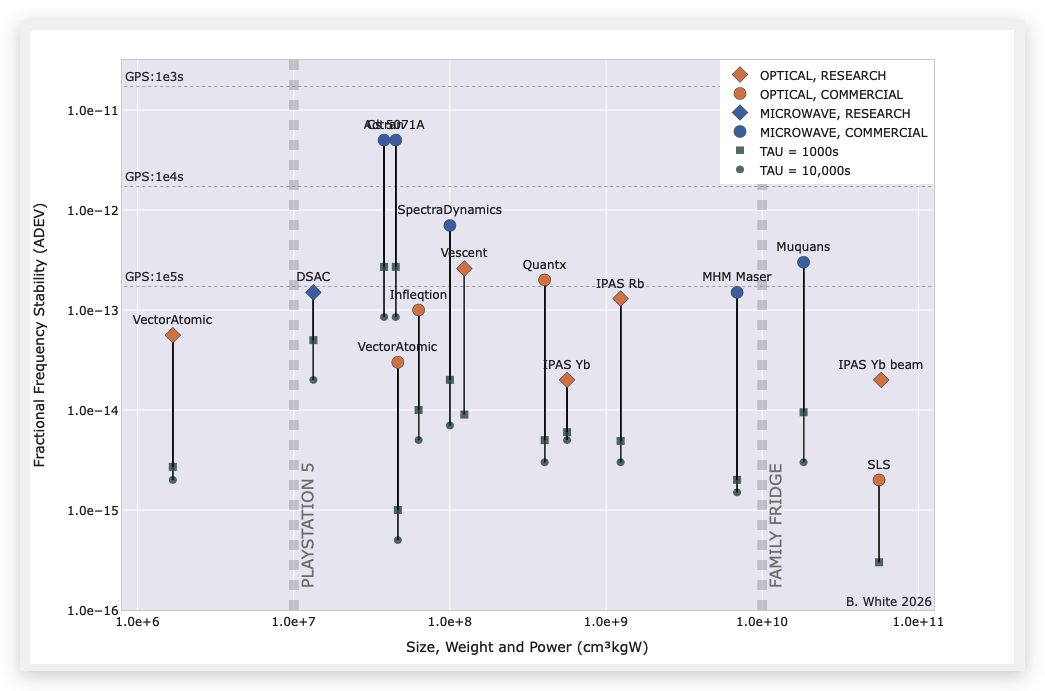}
    \caption{
        Example of using the zoom feature to focus on a specific region of the plot. Over the years the density of devices on the SWaP plot has increased to the point that the clustering of data points can make some labels difficult to discern. The viewer can easily focus on a specific region of interest by drawing a box around those values. Double-clicking anywhere on the plot will take the viewer back to the full plot.}
    \label{fig:FOCUS-SWAP-PLOT}
\end{figure}

\section{The SWaP Plot}
    The SWaP plot (figure \ref{fig:FULL-SWAP-PLOT}) shows the Allan Deviation (ADEV) of portable atomic clocks as a function of system SWaP in units of \qty{}{\cm^3\kg\W}. Frequency stability is shown for $10^0$, $10^3$ and $10^4$ second integration times (where available). The online plot has interactive elements, for example: hovering over a point will reveal more information about the clock (subject to change): 

    \begin{itemize}
        \item Whether the clock is optical or microwave 
        \item Whether the clock is research or commercial 
        \item The SWaP value
        \item The ADEV at 1 second
        \item The manufacturer/research group
        \item The atomic species
        \item Brief descriptor of physics package
        \item Reference details
    \end{itemize}

    The current version of the plot has a menu that will appear when hovering over the very top of the plot, allowing the user to download the latest plot as a static PNG. Two areas on the plot serve as reference for performance: the dashed lines indicative of GNSS timing expectations, and the shaded region labeled: Lab Clock Performance, referring to the performance of the current gold standard of lab-based optical atomic clocks \cite{Schioppo2017}. 

    Many clocks are clustered in the central region of the SWaP plot. We expect that the density of clocks in this region of the plot will increase as more devices are produced. This clustering makes the formatting of legible labels of each clock difficult. Figure \ref{fig:FOCUS-SWAP-PLOT} is an example of how the viewer can zoom in on a region of the plot, which allows individual devices to more easily be resolved and compared. All interactive features work while zoomed in, including the ability to download a PNG of the current view.

\section{Links}
\label{sec:iLinks}
    The SWaP plot will be available at:
    
    \href{https://adelaide.edu.au/research/institute-for-photonics-advanced-sensing-and-quantum-technologies/research-capabilities/precision-measurement-group/}{https://adelaide.edu.au/research/institute-for-photonics-advanced-sensing-and-quantum-technologies/research-capabilities/precision-measurement-group/}.
    
    Further information, such as inclusion criteria for clocks and definitions, other variations of the SWaP plot, and all source code and data will be available at:
    
    \href{https://a1120960.github.io/PAC-SWaP/}{https://a1120960.github.io/PAC-SWaP/} 
    
    and the associated GitHub repository:
    
    \href{https://github.com/a1120960/PAC-SWaP}{https://github.com/a1120960/PAC-SWaP}.

\section{Goals}
    The goal of this website is to provide a constantly updated resource for anyone interested in the field of precision timing and its applications. We will endeavor to keep all information up to date as new devices are published or released onto the market, as well as updating the source data if and when any existing devices improve, as that information becomes available. We would encourage the community to contact us via the email address at the top of the paper if they would like to include additional portable atomic clocks on the website or give other feedback.

    We are publishing this work under a Creative Commons (CC-BY-SA-4.0) license, meaning that we are happy for the work to be used as long as it is cited via the DOI link for this paper. 

    We provide the reference list for those research clocks on the current SWaP plot (figure \ref{fig:FULL-SWAP-PLOT}). However, as this resource is tracking an ever-evolving field, it will always be best to refer to the website for the latest version.

\section{Acknowledgment}
This research is supported by the
Commonwealth of Australia as represented by the Defence Science and Technology Group of the Department of Defence.

\bibliographystyle{ieeetr}
\bibliography{references}

@article{Takamoto2020a,
author = {Takamoto, Masao and Ushijima, Ichiro and Ohmae, Noriaki and Yahagi, Toshihiro and Kokado, Kensuke and Shinkai, Hisaaki and Katori, Hidetoshi},
doi = {10.1038/s41566-020-0619-8},
issn = {1749-4885},
journal = {Nature Photonics},
number = {7},
pages = {411--415},
publisher = {Springer US},
title = {{Test of general relativity by a pair of transportable optical lattice clocks}},
url = {http://dx.doi.org/10.1038/s41566-020-0619-8 https://www.nature.com/articles/s41566-020-0619-8},
volume = {14},
year = {2020}
}

@article{Burt2021,
author = {Burt, E. A. and Prestage, J. D. and Tjoelker, R. L. and Enzer, D. G. and Kuang, D. and Murphy, D. W. and Robison, D. E. and Seubert, J. M. and Wang, R. T. and Ely, T. A.},
doi = {10.1038/s41586-021-03571-7},
issn = {0028-0836},
journal = {Nature},
number = {7865},
pages = {43--47},
title = {{Demonstration of a trapped-ion atomic clock in space}},
url = {http://feeds.nature.com/$\sim$r/nature/rss/current/$\sim$3/Mxilx1ekfmM/s41586-021-03571-7?utm_source=researcher_app&utm_medium=referral&utm_campaign=RESR_MRKT_Researcher_inbound http://www.nature.com/articles/s41586-021-03571-7},
volume = {595},
year = {2021}
}

@article{Huang2020,
author = {Huang, Yao and Zhang, Huaqing and Zhang, Baolin and Hao, Yanmei and Guan, Hua and Zeng, Mengyan and Chen, Qunfeng and Lin, Yige and Wang, Yuzhuo and Cao, Shiying and Liang, Kun and Fang, Fang and Fang, Zhanjun and Li, Tianchu and Gao, Kelin},
doi = {10.1103/PhysRevA.102.050802},
issn = {24699934},
journal = {Physical Review A},
number = {5},
pages = {050802},
publisher = {American Physical Society},
title = {{Geopotential measurement with a robust, transportable Ca+ optical clock}},
url = {https://link.aps.org/doi/10.1103/PhysRevA.102.050802},
volume = {102},
year = {2020}
}

@article{Koller2017,
author = {Koller, S. B. and Grotti, J. and Vogt, St. and Al-Masoudi, A. and D{\"{o}}rscher, S. and H{\"{a}}fner, S. and Sterr, U. and Lisdat, Ch.},
doi = {10.1103/PhysRevLett.118.073601},
issn = {1079-7114},
journal = {Physical review letters},
number = {7},
pages = {073601},
pmid = {28256845},
publisher = {American Physical Society},
title = {{Transportable Optical Lattice Clock with 7×10$^{-17}$ Uncertainty.}},
url = {https://link.aps.org/doi/10.1103/PhysRevLett.118.073601 http://www.ncbi.nlm.nih.gov/pubmed/28256845},
volume = {118},
year = {2017}
}

@article{Poli2014,
author = {Poli, N. and Schioppo, M. and Vogt, S. and Falke, St and Sterr, U. and Lisdat, Ch and Tino, G. M.},
doi = {10.1007/s00340-014-5932-9},
issn = {0946-2171},
journal = {Applied Physics B},
number = {4},
pages = {1107--1116},
title = {{A transportable strontium optical lattice clock}},
url = {http://arxiv.org/abs/1409.4572 http://dx.doi.org/10.1007/s00340-014-5932-9 http://link.springer.com/10.1007/s00340-014-5932-9},
volume = {117},
year = {2014}
}

@article{Schioppo2017,
author = {Schioppo, M. and Brown, R. C. and McGrew, W. F. and Hinkley, N. and Fasano, R. J. and Beloy, K. and Yoon, T. H. and Milani, G. and Nicolodi, D. and Sherman, J. A. and Phillips, N. B. and Oates, C. W. and Ludlow, A. D.},
doi = {10.1038/nphoton.2016.231},
issn = {17494893},
journal = {Nature Photonics},
number = {1},
pages = {48--52},
publisher = {Nature Publishing Group},
title = {{Ultrastable optical clock with two cold-atom ensembles}},
volume = {11},
year = {2017}
}

@article{Khabarova2022,
author = {Khabarova, Ksenia and Kryuchkov, Denis and Borisenko, Alexander and Zalivako, Ilia and Semerikov, Ilya and Aksenov, Mikhail and Sherstov, Ivan and Abbasov, Timur and Tausenev, Anton and Kolachevsky, Nikolay},
doi = {10.3390/sym14102213},
issn = {20738994},
journal = {Symmetry},
number = {10},
pages = {1--15},
title = {{Toward a New Generation of Compact Transportable Yb+ Optical Clocks}},
volume = {14},
year = {2022}
}

@misc{CSAC2024,
  author       = {Microsemi},
  title        = {{Microsemi-CSAC-SA.45}},
  howpublished = {\url{https://www.microsemi.com/product-directory/embedded-clocks-frequency-references/5207-space-csac}},
  year         = {2024},
  note         = {Accessed: July 18, 2024}
}

@misc{PRS102024,
  author       = {Stanford Research Systems},
  title        = {{SRS-PRS10}},
  howpublished = {\url{https://www.thinksrs.com/products/prs10.html}},
  year         = {2024},
  note         = {Accessed: July 18, 2024}
}

@misc{5071A2024,
  author       = {Microchip Technology},
  title        = {{Microchip-5071A}},
  howpublished = {\url{https://www.microchip.com/en-us/products/clock-and-timing/components/atomic-clocks/atomic-system-clocks/cesium-time/5071a}},
  year         = {2024},
  note         = {Accessed: July 18, 2024}
}

@misc{MHMMaser2024,
  author       = {Microchip Technology},
  title        = {{Mirochip MHM-2020 Maser}},
  howpublished = {\url{https://www.microchip.com/en-us/products/clock-and-timing/components/atomic-clocks/atomic-system-clocks/mhm-2020-hydrogen-masers}},
  year         = {2024},
  note         = {Accessed: July 18, 2024}
}

@misc{NPLCsFountain2024,
  author       = {National Physical Laboratory},
  title        = {{NPL caesium Fountain}},
  howpublished = {\url{https://www.npl.co.uk/instruments/caesium-fountain}},
  year         = {2024},
  note         = {Accessed: July 18, 2024}
}

@misc{MUCLOCK2024,
  author       = {Muquans},
  title        = {{Muquans MUCLOCK}},
  howpublished = {\url{https://www.muquans.com/product/muclock/}},
  year         = {2024},
  note         = {Accessed: July 18, 2024}
}

@misc{cRb2024,
  author       = {SpectraDynamics},
  title        = {{SpectraDymanics cRb}},
  howpublished = {\url{https://spectradynamics.com/products/crb-clock/}},
  year         = {2024},
  note         = {Accessed: July 18, 2024}
}

@misc{VECTORATOMIC2024,
  author       = {Vector Atomic},
  title        = {{Vector Atomic EG-30}},
  howpublished = {\url{https://vectoratomic.com}},
  year         = {2024},
  note         = {Accessed: July 18, 2024}
}

@misc{TIQKER2024,
  author       = {Infleqtion},
  title        = {{Infleqtion TIQKER}},
  howpublished = {\url{https://www.infleqtion.com/tiqker}},
  year         = {2024},
  note         = {Accessed: July 18, 2024}
}

@misc{OPTICLOCK2024,
  author       = {BMBF},
  title        = {{BMBF-OPTICLOCK}},
  howpublished = {\url{https://www.opticlock.de/en/info}},
  year         = {2024},
  note         = {Accessed: July 18, 2024}
}

@misc{DFMVESCENT2024,
  author       = {DFM and Vescent Photonics},
  title        = {{DFM/VESCENT acetylene clock}},
  howpublished = {\url{https://vescent.com/media/wysiwyg/Products/FFC-Stabilaser_White_Paper.pdf}},
  year         = {2024},
  note         = {Accessed: July 18, 2024}
}

@article{Bothwell2025,
  author = {Bothwell, Tobias and Brand, Wesley and Fasano, Robert and Akin, Thomas and Whalen, Joseph and Grogan, Tanner and Chen, Yun-Jhih and Pomponio, Marco and Nakamura, Takuma and Rauf, Benjamin and Baldoni, Ignacio and Giunta, Michele and Holzwarth, Ronald and Nelson, Craig and Hati, Archita and Quinlan, Franklyn and Fox, Richard and Peil, Steven and Ludlow, Andrew},
  doi = {10.1364/OL.543310},
  issn = {0146-9592},
  journal = {Optics Letters},
  number = {2},
  pages = {646},
  pmid = {39815583},
  publisher = {Optica Publishing Group},
  title = {{Deployment of a transportable Yb optical lattice clock}},
  url = {http://dx.doi.org/10.1364/OL.543310 https://opg.optica.org/abstract.cfm?URI=ol-50-2-646},
  volume = {50},
  year = {2025},
}

@article{Offer2026,
  author = {Offer, R. F. and Klantsataya, E. and Hilton, A. P. and Strathearn, A. and {Bourbeau H{\'{e}}bert}, N. and Billington, C. J. and Watzdorf, S. and Scholten, S. K. and White, B. and Nelligan, M. and Stace, T. M. and Luiten, A. N.},
  title = {{Portable laser-cooled ytterbium beam clock based on an ultra-narrow optical transition}},
  journal = {Optica},
  year = {2026},
  doi = {10.1364/OPTICA.584095},
  issn = {2334-2536},
  number = {4},
  pages = {677},
  url = {http://arxiv.org/abs/2603.25261 https://opg.optica.org/abstract.cfm?URI=optica-13-4-677},
  volume = {13},
}

@article{Hilton2025,
  author = {Hilton, A. P. and Offer, R. F. and Klantsataya, E. and Scholten, S. K. and {Bourbeau H{\'{e}}bert}, N. and Billington, C. J. and Locke, C. and Perrella, C. and Nelligan, M. and Allison, J. W. and White, B. and Ahern, E. and Martin, K. W. and Beard, R. and Elgin, J. D. and Sparkes, B. M. and Luiten, A. N.},
  title = {{Demonstration of a mobile optical clock ensemble at sea}},
  journal = {Nature Communications},
  year = {2025},
  doi = {10.1038/s41467-025-61140-2},
  issn = {2041-1723},
  number = {1},
  pages = {6063},
  publisher = {Springer US},
  url = {https://www.nature.com/articles/s41467-025-61140-2},
  volume = {16},
}

@misc{AdtranOSA33002024,
  author       = {Adtran (Oscilloquartz)},
  title        = {{OSA-3300 Series}},
  howpublished = {\url{https://www.oscilloquartz.com/en/products-and-services/osa-3300-series}},
  year         = {2026},
  note         = {Accessed: September 19, 2025}
}

@misc{SLS2026,
  author       = {Stable Laser Systems},
  title        = {{Sr+ Ion Clock}},
  howpublished = {\url{https://stablelasers.com}},
  year         = {2026},
  note         = {Accessed: February 11, 2025. Clock measurements performed using the self-comparison technique described by Dub\'e et al., Phys. Rev. A 92, 042119 (2015), https://doi.org/10.1103/PhysRevA.92.042119}
}

\end{document}